\documentclass[11pt,letterpaper,oneside]{article}
\usepackage{graphicx}
\usepackage{amssymb}
\usepackage{amsmath}
\usepackage{bm}
\usepackage{setspace}
\usepackage{moreverb}
\usepackage{natbib}
\usepackage{amsfonts}
\usepackage{url}
\usepackage[usenames]{color}
\usepackage{listings}
\usepackage{courier}
\lstdefinelanguage{CUDA}{morekeywords={malloc,free},morekeywords={[2]cudaMalloc,cudaFree,cudaMemcpy,cudaMemcpyHostToDevice,__global__,__device__}}

\usepackage[letterpaper, hmargin=1in, vmargin=1in]{geometry}
\providecommand{\X}{\mathcal{X}}
\providecommand{\A}{\mathcal{A}}
\providecommand{\M}{\mathcal{M}}
\providecommand{\N}{\mathcal{N}}
\providecommand{\I}{\mathcal{I}}
\renewcommand{\Re}{\mathbb{R}}
\providecommand{\ind}{\mathbb{I}}
\providecommand{\bx}{{\bf x}}
\providecommand{\by}{{\bf y}}
\providecommand{\eqdef}{\triangleq}

\bibpunct{(}{)}{;}{a}{}{,}
\begin{document}

%\author{Anthony Lee \footnote{University of Oxford, Department of Statistics, 1 South Parks Road, Oxford OX1 3TG, UK, and Oxford-Man Institute, 9 Alfred Street, Oxford OX1 4EH, UK. Phone: +44 7595 713922. Fax: +44 1865 616601. Email: lee@stats.ox.ac.uk}
%\and Christopher Yau \footnote{University of Oxford, Department of Statistics, 1 South Parks Road, Oxford OX1 3TG, UK. Phone: +44 1865 285364. Fax: +44 1865 272595. Email: yau@stats.ox.ac.uk}
%\and Michael B. Giles \footnote{University of Oxford, Mathematical Institute, 24-29 St Giles, Oxford OX1 3LB, UK, and Oxford-Man Institute, 9 Alfred Street, Oxford OX1 4EH, UK. Phone +44 1865 283872. Fax: +44 1865 616601. Email: mike.giles@maths.ox.ac.uk}
%\and Arnaud Doucet \footnote{Institute of Statistical Mathematics, 4-6-7 Minami-Azabu, Minato-ku, Tokyo 106-8569, Japan. Phone: +81 3 3446 1501. Fax: +81 3 5421 8796. Email: arnaud@cs.ubc.ca}
%\and Christopher C. Holmes \footnote{University of Oxford, Department of Statistics, 1 South Parks Road, Oxford OX1 3TG, UK, and Oxford-Man Institute, 9 Alfred Street, Oxford OX1 4EH, UK. Phone: +44 1865 285368. Fax: +44 1865 616601. Email: cholmes@stats.ox.ac.uk}
%}
\author{Anthony Lee \footnote{Oxford-Man Institute, 9 Alfred Street, Oxford OX1 4EH, UK, and University of Oxford, Department of Statistics, 1 South Parks Road, Oxford OX1 3TG, UK. Email: lee@stats.ox.ac.uk}
\and Christopher Yau \footnote{University of Oxford, Department of Statistics, 1 South Parks Road, Oxford OX1 3TG, UK. Email: yau@stats.ox.ac.uk}
\and Michael B. Giles \footnote{University of Oxford, Mathematical Institute, 24-29 St Giles, Oxford OX1 3LB, UK, and Oxford-Man Institute, 9 Alfred Street, Oxford OX1 4EH, UK. Email: mike.giles@maths.ox.ac.uk}
\and Arnaud Doucet \footnote{Institute of Statistical Mathematics, 4-6-7 Minami-Azabu, Minato-ku, Tokyo 106-8569, Japan. Email: arnaud@cs.ubc.ca}
\and Christopher C. Holmes \footnote{University of Oxford, Department of Statistics, 1 South Parks Road, Oxford OX1 3TG, UK, and Oxford-Man Institute, 9 Alfred Street, Oxford OX1 4EH, UK. Email: cholmes@stats.ox.ac.uk}
}
\date{\today}
\title{On the utility of graphics cards to perform massively parallel simulation of advanced Monte Carlo methods}
\maketitle

\begin{abstract}
We present a case-study on the utility of graphics cards to perform massively parallel simulation of advanced Monte Carlo methods. Graphics cards, containing multiple Graphics Processing Units (GPUs), are self-contained parallel computational devices that can be housed in conventional desktop and laptop computers. For certain classes of Monte Carlo algorithms they offer massively parallel simulation, with the added advantage over conventional distributed multi-core processors that they are cheap, easily accessible, easy to maintain, easy to code, dedicated local devices with low power consumption. On a canonical set of stochastic simulation examples including population-based Markov chain Monte Carlo methods and Sequential Monte Carlo methods, we find speedups from 35 to 500 fold over conventional single-threaded computer code. Our findings suggest that GPUs have the potential to facilitate the growth of statistical modelling into complex data rich domains through the availability of cheap and accessible many-core computation. We believe the speedup we observe should motivate wider use of parallelizable simulation methods and greater methodological attention to their design.

\end{abstract}

{\small \textbf{Keywords:} Sequential Monte Carlo, Population-Based Markov Chain Monte Carlo, General Purpose Computation on Graphics Processing Units, Many-Core Architecture, Stochastic Simulation, Parallel Processing }

\section{Introduction}
We describe a case-study in the utility of graphics cards involving Graphics Processing Units (GPUs) to perform local, dedicated, massively parallel stochastic simulation. GPUs were originally developed as dedicated devices to aid in real-time graphics rendering. However recently there has been an emerging literature on their use for scientific computing as they house multi-core processors. Examples include \cite{mm} and \cite{mds}, which discuss their use in molecular modelling and dynamics.

To gain an understanding of the potential benefits to statisticians we have investigated speedups on a canonical set of examples taken from the advanced Monte Carlo literature. These include Bayesian inference for a Gaussian mixture model computed using a population-based Markov chain Monte Carlo (MCMC) method and a sequential Monte Carlo (SMC) sampler and sequential Bayesian inference for a multivariate stochastic volatility model implemented using a standard SMC method, also known as a particle filter in this context. In these examples we report substantial speedups from the use of GPUs over conventional CPUs.

The potential of parallel processing to aid in statistical computing is well documented (see e.g. \cite{hpcs}). However, previous studies have relied on distributed multi-core clusters of CPUs for implementation. In contrast, graphics cards for certain generic types of computation offer parallel processing speedups with advantages on a number of fronts, including:

\begin{itemize}
\item Cost: graphics cards are relatively cheap. At time of writing the cards used in our study retail at around 200 US dollars.

\item Accessibility: graphics cards are readily obtainable from high street computer stores or over the internet.

\item Maintenance: the devices are self-contained and can be hosted on conventional desktop and laptop computers.

\item Speed: in line with multi-core CPU clusters, graphics cards offer significant speedup, albeit for a restricted class of scientific computing algorithms.

\item Power: GPUs are low energy consumption devices compared to clusters of traditional computers, with a graphics card requiring around 200 Watts. While improvements in energy efficiency are application-specific, it is reasonable in many situations to expect a GPU to use around 10 per cent of the energy to that of an equivalent CPU cluster.

\item Dedicated and local: the graphics cards slot into conventional computers offering the user ownership without the need to transport data externally.
\end{itemize}

The idea of splitting the computational effort of parallelizable algorithms amongst processors is certainly not new to statisticians. In fact, distributed systems and clusters of computers have been around for decades. Previous work on parallelization of MCMC methods on a group of networked computers include, among others, \cite{rosenthal} and \cite{brockwell}. \cite{rosenthal} discusses how to deal with computers running at different speeds and potential computer failure while \cite{brockwell} discusses the parallel implementation of a standard single chain MCMC algorithm by pre-computing acceptance ratios. The latency and bandwidth of communication in these systems make them suitable only in cases where communication between streams of computation, or threads, is infrequent and low in volume. In other words, while many algorithms involve computation that could theoretically be distributed amongst processors, the overhead associated with distributing the work erases any speedup. In contrast, many-core processor communication has very low latency and very high bandwidth due to high-speed memory that is shared amongst the cores. Low latency here means the time for a unit of data to be accessed or written to memory by a processor is low whilst high bandwidth means that the amount of data that can be sent in a unit of time is high. For many algorithms, this makes parallelization viable where it previously was not. In addition, the energy efficiency of a many-core computation compared to a single-core or distributed computation can be improved. This is because the computation can both take less time and require less overhead.

We choose to investigate the speed up for the simulation of random variates from complex distributions, a common computational task when performing inference using Monte Carlo (see e.g. \cite{mcsm}). In particular, we focus on population-based MCMC methods and SMC methods for producing random variates as these are not algorithms that typically see significant speedup on clusters due to the need for frequent, high-volume communication between computing nodes. For the examples we consider, we find that computation time can be significantly lowered for all applications, and drastically lowered in some cases. This means that we can obtain the samples we want in seconds instead of hours and minutes instead of days. The types of speedup observed are dependent on the ability of the methods to be parallelized. In particular, speedup increases with the number of auxiliary distributions in population-based MCMC and the number of particles in SMC until some device-specific capacity is reached.

The algorithms are implemented for the Compute Unified Device Architecture (CUDA) and make use of GPUs which support this architecture. CUDA offers a fairly mature development environment via an extension to the C programming language. We estimate that a programmer proficient in C should be able to code effectively in CUDA within a few weeks of dedicated study. For our applications we use CUDA version 2.1 with an NVIDIA GTX 280 as well as an NVIDIA 8800 GT. The GTX 280 has 30 multiprocessors while the 8800 GT has 14 multiprocessors. For all current NVIDIA cards, a multiprocessor comprises 8 arithmetic logic units (ALUs), 2 special units for transcendental functions, a multithreaded instruction unit and on-chip shared memory. In addition to having more multiprocessors than the 8800 GT, each GTX 280 multiprocessor itself has more registers, can support more active threads and includes one double-precision ALU. For single-precision floating point computation, one can think of the GTX 280 as having 240 ($30 \times 8$) and the 8800 GT as having 112 ($14 \times 8$) single processors respectively. At present, the retail price of the GTX 280 is just over double that of the 8800GT and it requires just over twice the power.

\section{GPUs for Parallel Processing}
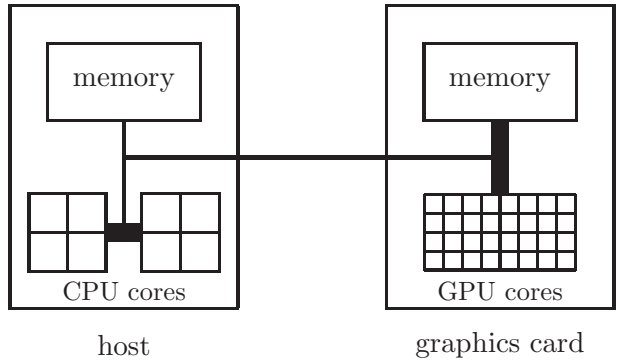
\begin{figure}[htp]
\center
{\setlength{\unitlength}{1.0cm}
\begin{center}\begin{picture}(8,5)(0,-1)
\thicklines
\put(0.0,0.0){\framebox(3,4){}}
\put(0.5,2.5){\framebox(2,1){memory}}

\multiput(1.49,1.0)(0.01,0){3}{\line(0,1){1.5}}
\multiput(1.25,0.9)(0,0.01){21}{\line(1,0){0.5}}

\multiput(0.25,0.5)(1.5,0){2}{\framebox(1,1){}}
\multiput(0.25,1.0)(1.5,0){2}{\line(1,0){1}}
\multiput(0.75,0.5)(1.5,0){2}{\line(0,1){1}}

\multiput(1.5,1.99)(0,0.01){3}{\line(1,0){5}}

\put(5.0,0.0){\framebox(3,4){}}
\put(5.5,2.5){\framebox(2,1){memory}}

\multiput(6.4,1.5)(0.01,0){21}{\line(0,1){1}}

\multiput(5.5,0.5)(0,0.25){5}{\line(1,0){2}}
\multiput(5.5,0.5)(0.25,0){9}{\line(0,1){1}}

\put(1.5,0.22){\makebox(0,0){\small CPU cores}}
\put(6.5,0.22){\makebox(0,0){\small GPU cores}}

\put(1.5,-0.5){\makebox(0,0){host}}
\put(6.5,-0.5){\makebox(0,0){graphics card}}

\end{picture}\end{center}}
\caption[Link between host and graphics card]{Link between host and graphics card. The thicker lines represent higher data bandwidth while the squares represent processor cores.}
\label{fig:gpu_cpu_link}
\end{figure}

GPUs have evolved into many-core processing units, currently with up to 30 multiprocessors per card, in response to commercial demand for real-time graphics rendering, independently of demand for many-core processors in the scientific computing community. As such, the architecture of GPUs is very different to that of conventional central processing units (CPUs). An important difference is that GPUs devote proportionally more transistors to ALUs and less to caches and flow control in comparison to CPUs. This makes them less general purpose but highly effective for data-parallel computation with high arithmetic intensity, i.e. computations where the same instructions are executed on different data elements and where the ratio of arithmetic operations to memory operations is high. This single instruction, multiple data (SIMD) architecture puts a heavy restriction on the types of computation that optimally utilize the GPU but in cases where the architecture is suitable it reduces overhead.

Figure \ref{fig:gpu_cpu_link} gives a visualization of the link between a host machine and the graphics card, emphasizing the data bandwidth characteristics of the links and the number of processing cores. A program utilizing a GPU is hosted on a CPU with both the CPU and the GPU having their own memory. Data is passed between the host and the device via a standard memory bus, similar to how data is passed between main memory and the CPU. The memory bus between GPU memory and the GPU cores is both wider and has a higher clock rate than a standard bus, enabling much more data to be sent to the cores than the equivalent link the host. This type of architecture is ideally suited to data-parallel computation since large quantities of data can be loaded into registers for the cores to process in parallel. In contrast, typical computer architectures use a cache to speed up memory accesses using locality principles that are generally good but do not fully apply to data-parallel computations, with the absence of temporal locality most notable.

\subsection{Programming with Graphics Cards}
\label{section:programming_gpus}
CUDA provides the interface to compliant GPUs by extending the C programming language. Programs compiled with CUDA allow computation to be split between the CPU and the GPU. In this sense, the GPU can be treated as an additional, specialized processor for data-parallel computation. In the following text, host code refers to code that is executed on the CPU whilst device code is code that is executed on the GPU. We present a simple example in Figures \ref{code:is_global}-\ref{code:is_host}, explained below, that computes a classical importance sampling estimate (see Section \ref{section:methods}). In the code snippets, keywords in the C language are in bold face whilst CUDA keywords are both bold and italicized. A line beginning with a ``{\tt //}'' is a comment and is ignored by the compiler.

\begin{figure}
\begin{lstlisting}[frame=single]
__global__ void importance_sample(int N, float* d_array, float* d_array_out) {
    // thread id = threads per block * block id + thread id within block
    const int tid = blockDim.x * blockIdx.x + threadIdx.x;
    // total number of threads = threads per block * number of blocks
    const int tt = blockDim.x * gridDim.x;
    int i;
    float w, x;
    for (i = tid; i < N; i += tt) {
        x = d_array[i];
        w = target_pdf(x) / proposal_pdf(x);
        d_array_out[i] = phi(x) * w;
    }
}
\end{lstlisting}
\caption{Kernel that evaluates an importance weight and test function}
\label{code:is_global}
\end{figure}

CUDA allows users to define special functions, called kernels, that are called by the host code to be executed in parallel on the GPU by a collection of threads. Figure \ref{code:is_global} shows an example of a kernel function, which can be invoked in host code using the syntax
\begin{verbatim}
    importance_sample<<<nb,nt>>>(N, d_array, d_array_out);
\end{verbatim}
where {\tt nb} is the number of blocks of threads and {\tt nt} is the number of threads per block. The total number of threads created by this call is the product of {\tt nb} and {\tt nt} and one can think of a threads as being a single stream of computation. For most kernels, the numbers of threads and blocks can be changed to tune performance on different cards or with different data. A more detailed description of blocks and threads and their relation to the hardware is given in Section \ref{section:blocks_threads}.

\begin{figure}
\begin{lstlisting}[frame=single]
__device__ float target_pdf(float x) {
    return 1.0f / sqrtf(2 * PI) * exp(-(x - 1.5) * (x - 1.5) / 0.5f)
         + 1.0f / sqrtf(2 * PI) * exp(-(x + 1)   * (x + 1)   / 0.5f);
}

__device__ float proposal_pdf(float x) {
    return 1.0f / sqrtf(2 * PI) * exp(-x * x / 2.0f);
}

__device__ float phi(float x) {
    return x * x;
}
\end{lstlisting}
\caption{Device functions for evaluating the target density, the proposal density and the test function. The target is an equally weighted, two-component mixture of normals with equal variances of 0.25 and means at -1 and 1.5 while the proposal is a standard normal distribution. The test function squares its input so that the integral that is estimated is the expectation of the second moment of a random variable distributed according to the target density.}
\label{code:is_device}
\end{figure}

A kernel is defined with the {\tt \_\_global\_\_} qualifier. Kernels are special in that they are always invoked in parallel with the numbers of blocks and threads specified and have a void return type. As such, program correctness depends only on how the threads invoked in the kernel call modify memory on the graphics card. In particular, code must be written that guarantees the correct modifications to memory once all threads have completed, especially when threads interact during execution by reading and writing from shared memory locations. In Figure \ref{code:is_global}, a kernel is defined that takes as input an array of random values sampled from a proposal distribution and places, for each value, the product of the test function and the importance weight at that value in a separate array. One can see that each thread is responsible for ${\tt N} / {\tt tt}$ values, assuming ${\tt N}$ is a multiple of {\tt tt}. Within a kernel, special functions can be called that have been defined with the {\tt \_\_device\_\_} qualifier. These functions can only be called by {\tt \_\_global\_\_} functions or {\tt \_\_device\_\_} functions themselves. In Figure \ref{code:is_global}, {\tt target\_pdf}, {\tt proposal\_pdf} and {\tt phi} are examples of this, and their definitions are provided in Figure \ref{code:is_device}. In this particular kernel we see that each thread first computes its absolute thread identifier {\tt tid} and the total number of threads {\tt tt}. It then computes an importance weight and evaluates the test function for each value in {\tt d\_array} it is responsible for and stores the result in {\tt d\_array\_out}. Since there is no thread interaction in this example kernel, it is reasonably straightforward to verify its correctness.

\begin{figure}
\begin{lstlisting}[frame=single]
int N = 16777216;

float h_sum, result;
float* d_array;
float* d_array_out;

float* array = (float*) malloc(N * sizeof(float));
cudaMalloc((void **) &d_array, N * sizeof(float));
cudaMalloc((void **) &d_array_out, N * sizeof(float));

populate_randn(array, N);

cudaMemcpy(d_array, array, N * sizeof(float), cudaMemcpyHostToDevice);

is<<<64,128>>>(N, d_array, d_array_out);
h_sum = reduce(N, d_array_out);
result = h_sum / N;

free(array);
cudaFree(d_array);
cudaFree(d_array_out);
\end{lstlisting}
\caption{Host code}
\label{code:is_host}
\end{figure}

Figure \ref{code:is_host} gives a snippet of code that is run on the host and completes our example. First, memory is allocated on both the host and the graphics card using the {\tt malloc} and {\tt cudaMalloc} functions respectively. The host function {\tt populate\_randn} then puts {\tt N} standard normal random variates in {\tt array}. These values are copied into the GPU array, {\tt d\_array}, via the {\tt cudaMemcpy} function. In Figure \ref{fig:gpu_cpu_link}, this is a transfer along the memory bus that connects host and graphics card memory. At this point, the kernel is called with 64 blocks of 128 threads per block. The {\tt reduce} function is a CPU function that returns the sum of the elements in a GPU array. Of course, this function can itself invoke a GPU kernel. Finally, the importance sampling estimate is obtained by dividing this sum by {\tt N} and memory is freed. Note that this code has been written so as to expose the most common functions that are used in GPU programming using CUDA. For example, it would be faster to create the random variates on the GPU itself but this would not have allowed any memory transfer operations to be shown here.

This basic example highlights the most important characteristics of CUDA programs: memory management, kernel specification and kernel invocation. Memory management is a key component in algorithm design using graphics cards since there is often need for transfer between CPU and GPU memory as standard host functions can only access CPU memory and kernels can only access GPU memory. The fundamental memory operations are {\tt cudaMalloc}, {\tt cudaMemcpy} and {\tt cudaFree}, which are GPU analogues to the standard C functions {\tt malloc}, {\tt memcpy} and {\tt free}. Kernel specification requires ensuring that correct output will be given once all threads have returned. In the above example, it is clear that all concurrent threads will be executing the same instructions in parallel because there is no conditional branching, which occurs when different instructions are executed in concurrent threads based on the result of a data-dependent runtime comparison. Indeed, while it is possible to specify arbitrary conditional branches within a kernel, this can lead to slow performance since threads in a SIMD architecture execute sequentially when they are not executing the same instructions, which can be devastating to performance. An important constraint on kernel code that is not illustrated explicitly in the above code is that neither recursive functions nor function pointers can be defined in device code. This is due to the fact that kernel functions are completely determined at compile time with {\tt \_\_device\_\_} functions simply inlined, or inserted into the kernel, during compilation. With respect to kernel invocation, the number of threads and blocks assigned to each kernel can be decided at runtime in host code. This is useful since computation time can depend strongly on these numbers and optimal configurations will vary across graphics cards and features of the data. A final remark is that the level of abstraction provided by CUDA is close to the hardware operations on the device. This ensures that programmers are acutely aware of the benefits of writing, for example, kernels with minimal interaction between threads and avoiding branching.

\subsection{Blocks and Threads}
\label{section:blocks_threads}
CUDA abstracts the hardware of the GPU into blocks and threads to simultaneously provide a relatively simple view of the architecture to developers while still allowing a low-level abstraction of the hardware for performance reasons. One can generally think of each thread as being computed on a virtual processor. The block abstraction is necessary to provide the concept of a virtual microprocessor. Threads within a block are capable of more interaction than threads in separate blocks, mainly due to the fact that all threads in a block will be executed on the same microprocessor. As such, they have access to very fast, dynamically allocated, on-chip memory and can perform simple barrier synchronization. In Section \ref{section:programming_gpus}, this advanced functionality is not required by the example kernel.

It is important to note that blocks and threads are still very much virtual constructs. At runtime, multiple blocks may be executed concurrently on the same multiprocessor. With respect to ALU execution, operations are performed on groups of 32 threads at a time, which allows each of the 8 scalar processors to perform 4 identical instructions in quick succession in an ideal setting. The group of 32 threads that executes simultaneously is called a warp. 

\subsection{Single Precision Issues}
The current generation of GPUs is 4-8 times faster at single precision arithmetic than double precision. Although this ratio will decrease in the future, there will probably remain a factor 2 difference in speed, the same as for Intel CPUs when using SSE instructions. This raises the question of whether single precision arithmetic is adequate for statistical applications.

There are two particular areas in which care must be taken. The first concerns the much more limited range of single precision floating point numbers. Because of their 8-bit exponent, their magnitude must lie in the approximate range $[10^{-38}, 10^{+38}]$, whereas the magnitude of double precision variables is in the approximate range $[10^{-308}, 10^{+308}]$. Consequently, when working in single precision it is often necessary to work with log-likelihoods, rather than the likelihoods themselves though this is rarely a restriction for statisticians.

The second area of potential issues concerns the averaging of $N$ floating point values, for $N\!\gg\! 1$. The simplest implementation uses an accumulator, to which the values are added one at a time. However, this may lead to a large increase in the error due to finite machine precision. When all of the values are of the same sign, the relative error is amplified by factor $O(N)$ in the worst case, and $O(\sqrt{N})$ in the more typical case where the rounding error at each step can be modelled as a random variable with zero mean. This behaviour is well understood \cite{higham93, higham02} and the growth can be reduced to $O(\log N)$ by using a binary tree summation algorithm in which the values are summed in pairs, and then those new values are summed in pairs, and the process is repeated until a single value is obtained. This is the natural approach for the parallel implementation of a reduction operation. Example code is provided by NVIDIA on their CUDA website, and the implementation in our code is based on this.

Despite these concerns, single precision seems perfectly sufficient for the applications in this paper. The statistical variability due to the use of random numbers within the algorithms exceeds the perturbations due to finite machine precision.

\subsection{GPU Parallelizable Algorithms}
In general, if a computing task is well-suited to SIMD parallelization then it will be well-suited to computation on a GPU. In particular, data-parallel computations with high arithmetic intensity (computations where where the ratio of arithmetic operations to memory operations is high) are able to attain maximum performance from a GPU. This is because the volume of very fast arithmetic instruction can hide the relatively slow memory accesses. It is crucial to determine whether a particular computation is data-parallel on the instruction level when determining suitability. From a statistical simulation perspective, integration via classical Monte Carlo or importance sampling are ideal computational tasks in a SIMD framework. This is because each computing node can produce and weight a sample in parallel, assuming that the sampling procedure and the weighting procedure have no conditional branches. If these methods do branch, speedup can be compromised by many computing nodes running idle while others finish their tasks. This can occur, for example, if the sampling procedure uses rejection sampling.

In contrast, if a computing task is not well-suited to SIMD parallelization then it will not be well-suited to computation on a GPU. In particular, task-parallel computations where one executes different instructions on the same or different data cannot utilize the shared flow control hardware on a GPU and often end up running sequentially. Even when a computation is data-parallel, it might not give large performance improvements on a GPU due to memory constraints. This can be due to the number of registers required by each thread (see Sections \ref{section:fsv} and \ref{section:discussion}) or due to the size and structure of the data necessary for the computation requiring large amounts of memory to be transferred between the host and the graphics card. The latter issue is analogous to the issue of thrashing in virtual memory systems and can occur, for example, when an algorithm iterates over a block of data that will not fit in memory.

There are also many computational tasks in statistical computing that are just difficult to parallelize. For example, standard Metropolis-Hastings MCMC with a single chain is difficult to parallelize in the general case because it is a naturally sequential algorithm. Parallelization of this type of method usually involves parallelization of the target density evaluation as in \cite{suchard_rambaut}, the sampling from or evaluation of the proposal density or computation of multiple possible execution paths as in \cite{brockwell} as opposed to parallelization of the general algorithm itself.

The availability of new hardware suited to parallel computation motivates use of a new model of computation for developing and analyzing the efficacy of statistical algorithms. In some cases, existing algorithms will require little modification to take advantage of this technology whilst in others major changes will have to be made. There is also the potential for previously impractical and novel algorithms to become important tools for statisticians.

\subsection{Parallel Random Number Generation}
One important aspect of any Monte Carlo simulation is the generation of pseudorandom numbers. Fortunately, many uniform pseudorandom number generators can be implemented efficiently in parallel. The key idea is that each thread computes a contiguous block of numbers within a single overall stream. The thread can jump to the start of its block of numbers using a ``skip-ahead'' algorithm which enables it to skip $n$ places in $O(\log n)$ operations (e.g. see \cite{esck02}). The uniform pseudorandom numbers can then be transformed to match various different output distributions as needed. In our applications we use a parallelized version of the multiple recursive generator MRG32k3a presented in \cite{ecuyer} as well as a parallelized version of a xorshift random number generator \cite{xorshift}. In the case of the xorshift random number generator, more time must be spent to compute the seeds for each thread before any computation is done but the random number generation itself is faster and the initialization can be done offline.

\section{Parallelizable Sampling Methods}
\label{section:methods}
In this section we consider a number of sampling methods for which parallel implementations can be produced without significant modification. There is an abundance of statistical problems that are essentially computational in nature, especially in Bayesian inference. In many such cases, the problem can be distilled into one of sampling from a probability distribution whose density $\pi$ we can compute pointwise and up to a normalizing constant, i.e. we can compute $\pi^{\ast}(\cdot)$ where $\pi(\bx)=\pi^{\ast}(\bx)/Z$. A common motivation for wanting samples from $\pi$ is so we can compute expectations of certain functions. If we denote such a function by $\phi$, the expectation of interest is
\[
I\eqdef\int_{\bx\in\X}\phi(\bx)\pi(\bx)d\bx
\]
The Monte Carlo estimate of this quantity is given by
\[
\hat{I}_{MC}\eqdef\frac{1}{N}\sum_{i=1}^{N}\phi(\bx^{(i)})
\]
where $\{\bx^{(i)}\}_{i=1}^{N}$ are samples from $\pi$.

Clearly, we need samples from $\pi$ in order to compute this estimate. In practice, however, we often cannot sample from $\pi$ directly. There are two general classes of methods for dealing with this. The first are importance sampling methods, where we generate weighted samples from $\pi$ by generating $N$ samples according to some importance density $\gamma$ proportional to $\gamma^{\ast}$ and then estimating $I$ via
\[
\hat{I}_{IS}\eqdef\sum_{i=1}^{N}W^{(i)}\phi(\bx^{(i)})
\]
where $W^{(i)}$ are normalized importance weights
\[
W^{(i)}=\frac{w(\bx^{(i)})}{\sum_{j=1}^{N}w(\bx^{(j)})}\text{ and } w(\bx^{(i)})=\frac{\pi^{\ast}(\bx^{(i)})}{\gamma^{\ast}(\bx^{(i)})}
\]
The asymptotic variance of this estimate is given by $C(\phi,\pi,\gamma)/N$, i.e. a constant over $N$. For many problems, however, it is difficult to come up with an importance density $\gamma$ such that $C(\phi,\pi,\gamma)$ is small enough for us to attain reasonable variance with practical values of $N$.

The second general class of methods are MCMC methods, in which we construct an ergodic $\pi$-stationary Markov chain sequentially. Once the chain has converged, we can use all the dependent samples to estimate $I$. The major issue with MCMC methods is that their convergence rate can be prohibitively slow in some applications.

There are many ways to parallelize sampling methods that are not the focus of this work. For example, naive importance sampling, like classical Monte Carlo, is intrinsically parallel. Therefore, in applications where we have access to a good importance density $\gamma$ we can get linear speedup with the number of processors available. Similarly, in cases where MCMC converges rapidly we can parallelize the estimation of $I$ by running separate chains on each processor. While these situations are hoped for, they are not particularly interesting from a parallel architecture standpoint because they can run equally well in a distributed system. Finally, this paper is not concerned with problems for which the computation of individual MCMC moves or importance weights are very expensive but themselves parallelizable. While the increased availability of parallel architectures will almost certainly be of help in such cases, the focus here is on potential speedups by parallelizing general sampling methods. An example of recent work in this area can be found in \cite{suchard_rambaut}, in which speedup is obtained by parallelizing evaluation of individual likelihoods.

Much work in recent years has gone into dealing with the large constants in the variance of importance sampling estimates and slow convergence rates in MCMC and it is in these `advanced' Monte Carlo methods that we direct our interest. This is mainly because while they are parallelizable, they are not trivially so and stand to benefit enormously from many-core architectures. In the remainder of this section we briefly review three such methods: population-based MCMC, SMC and SMC samplers.

\subsection{Population-Based MCMC}
A common technique in facilitating sampling from a complex distribution $\pi$ with support in $\X$ is to introduce an auxiliary variable ${\bf a} \in \A$ and sample from a higher-dimensional distribution $\bar{\pi}$ with support in the joint space $\A \times \X$, such that $\bar{\pi}$ admits $\pi$ as a marginal distribution. With such samples, one can discard the auxiliary variables and be left with samples from $\pi$.

This idea is utilized in population-based MCMC, which attempts to speed up convergence of an MCMC chain for $\pi$ by instead constructing a Markov chain on a joint space $\X^{M}$ using $M-1$ auxiliary variables each in $\X$. In general, we have $M$ parallel `subchains' each with stationary distribution $\pi_{i},i\in\M\eqdef\{1,\ldots,M\}$ and $\pi_{M}=\pi$. Associated with each subchain $i$ is an MCMC kernel $L_{i}$ that leaves $\pi_{i}$ invariant, and which we run at every time step. Of course, without any further moves, the stationary distribution of the joint chain is
\[
\bar{\pi}(\bx_{1:M})\eqdef\prod_{i=1}^{M}\pi_{i}(\bx_{i})
\]
and so $\bx_{M}\sim\pi$. This scheme does not affect the convergence rate of the independent chain $M$. However, since we can cycle mixtures of $\bar{\pi} $-stationary MCMC kernels without affecting the stationary distribution of the joint chain \cite{tierney}, we can allow certain types of interaction between the subchains which can speed up convergence \citep{geyer,hukushima_nemoto}. In general, we apply a series of kernels that act on subsets of the variables. For the sake of clarity, let us denote the number of second-stage kernels by $R$ and the kernels themselves as $K_{1},\ldots,K_{R}$, where kernel $K_{j}$ operates on variables with indices in $\I_{j}\subset\M$. The idea is that the $R$ kernels are executed sequentially and it is required that each $K_{j}$ leave $\prod_{i\in\I_{j}}\pi_{i}$ invariant.

Given $\pi$, there are a wide variety of possible choices for $M$, $\pi_{1:M-1}$, $L_{1:M}$, $R$, $\I_{1:R}$ and $K_{1:R}$ which will affect the convergence rate of the joint chain. For those interested, \cite{ajay} gives a review of some of these. It is clear that the first stage of moves involving $L_{1:M}$ is trivially parallelizable. However, the second stage is sequential in nature. For a parallel implementation, it is beneficial for the $\I_{j}$'s to be disjoint as this allows the sequence of exchange kernels to be run in parallel. Of course, this implies that $\I_{1:R}$ should vary with time since otherwise there will be no interaction between the disjoint subsets of chains. Furthermore, if the parallel architecture used is SIMD (Single Instruction Multiple Data) in nature, it is desirable to have the $K_{j}$'s be nearly identical algorithmically. The last consideration for parallelization is that while speedup is generally larger when more computational threads can be run in parallel, it is not always helpful to increase $M$ arbitrarily as this can affect the convergence rate of the chain. However, in situations where a suitable choice of $M$ is dwarfed by the number of computational threads available, one can always increase the number of chains with target $\pi$ to produce more samples.

\subsection{Sequential Monte Carlo}
SMC methods are a powerful extension of importance sampling methodology that are particularly popular for sampling from a sequence of probability distributions. In the context of state-space models, these methods are known as particle filtering methods; \cite{smc_tut} and \cite{liu} give recent surveys of the field. In this context, let $\left\{  \bx_{t}\right\}  _{t\geq0}$ be an unobserved Markov process of initial density $\bx_{0}\sim p_{0}(\cdot)$ and transition density $\bx_{t}\sim f(\cdot|\bx_{t-1})$ for $t\geq1$. We only have access to an observation process $\left\{  \by_{t}\right\}  _{t\geq0}$; the observations are conditionally independent conditional upon $\left\{\bx_{t}\right\}  _{t\geq0}$ of marginal density $\by_{t}\sim g(\by_{t} |\bx_{t})$ for $t\geq1$. SMC\ methods are used to approximate recursively in time the filtering densities $p(\bx_{0:t}|\by_{0:t})$ which are proportional to $p(\bx_{0:t},\by_{0:t})\eqdef p_{0}(\bx_{0})\prod_{i=1}^{t}f(\bx_{i} |\bx_{i-1})\prod_{i=1}^{t}g(\by_{i}|\bx_{i})$ for $t=1,\ldots,T$. These distributions are approximated with a set of random samples called particles through use of a sequential version of importance sampling and a special particle-interaction step known as resampling.

Parallelization of SMC methods is reasonably straightforward. The importance sampling step used at each time step is trivially parallelizable as it involves only the local state of a particle. The resampling step, in which some particles are replicated and others destroyed depending on their normalized importance weights, comprises the construction of an empirical cumulative distribution function for the particles based on their importance weights followed by sampling from this $N$ times, where $N$ is the fixed number of particles used throughout the computation. While neither of these tasks is trivially parallelizable, they can benefit moderately from parallelization. However, the bulk of the speedup will generally come from the parallelization of the evolution and weighting steps. Therefore, using criteria like effective sample size \citep{liu_chen} to avoid resampling at every time step can also improve speedup.

\subsection{Sequential Monte Carlo Samplers}
SMC samplers \citep{smcs} are a more general class of methods that utilize a sequence of auxiliary distributions $\pi_{0},\ldots,\pi_{T}$, much like population-based MCMC as discussed out in \cite{ajay}. However, in contrast to population-based MCMC, SMC samplers start from an auxiliary distribution $\pi_{0}$ and recursively approximate each intermediate distribution in turn until finally $\pi_{T}=\pi$ is approximated. The algorithm has the same general structure as classical SMC, with differences only in the types of proposal distributions, target distributions and weighting functions used in the algorithm. As such, parallelization of SMC samplers closely follows that of SMC.

The difference between population-based MCMC and SMC samplers is subtle but practically important. Both can be viewed as population-based methods on a similarly defined joint space since many samples are generated at each time step in parallel. However, in population-based MCMC the samples generated at each time each have different stationary distributions and the samples from a particular chain over time provide an empirical approximation of that chain's target distribution. In SMC samplers, the weighted samples generated at each time approximate one auxiliary target distribution and the true target distribution is approximated at the last time step. This difference is further discussed in Section \ref{section:comparison}.

\section{Canonical Examples}
To demonstrate the types of speed increase one can attain by utilizing GPUs, we apply each method to a representative statistical problem. We use Bayesian inference for a Gaussian mixture model as an application of the population-based MCMC and SMC samplers, while we use a factor stochastic volatility state-space model to gauge the speedup of our parallel SMC method. We ran our parallel code on a computer equipped with an NVIDIA 8800 GT GPU, a computer equipped with an NVIDIA GTX 280 GPU and we ran reference single-threaded code on a Xeon E5420 / 2.5 GHz processor. The resulting processing times and speedups are given in Tables \ref{tab:pop_mcmc} - \ref{tab:smc}.

The applications we discuss here are representative of the types of problems that these methods are commonly used to solve. In particular, while the distribution of mixture means given observations is only one example of a multimodal distribution, it can be thought of as a canonical distribution with multiple well-separated modes. Therefore, the ability to sample points from this distribution is indicative of the ability to sample points from a wide range of multimodal distributions. Similarly, performance of a latent variable sampler in dealing with observations from a factor stochastic volatility model is indicative of performance on observations from reasonably well-behaved but non-linear and non-Gaussian continuous state-space models.

\subsection{Mixture Modelling}
Finite mixture models are a very popular class of statistical models as they provide a flexible way to model heterogeneous data \citep{fmm}. Let $\by=y_{1:m}$ denote i.i.d. observations where $y_{j}\in\Re$ for $j\in\{1,\ldots,m\}$. A univariate Gaussian mixture model with $k$ components states that each observation is distributed according to the mixture density
\[
p(y_{j}|\mu_{1:k},\sigma_{1:k},w_{1:k-1})=\sum_{i=1}^{k}w_{i}f(y_{j}|\mu
_{i},\sigma_{i})
\]
where $f$ denotes the density of the univariate normal distribution. The density of $\by$ is then equal to $\prod_{j=1}^{m}p(y_{j}|\mu_{1:k}, \sigma_{1:k}, w_{1:k-1})$.

For simplicity, we assume that $k$, $w_{1:k-1}$ and $\sigma_{1:k}$ are known and that the prior distribution on $\mu$ is uniform on the $k$-dimensional hypercube $[-10,10]^{k}$. We set $k=4$, $\sigma_{i}=\sigma=0.55$, $w_{i}=w=1/k$ for $i\in\{1,\ldots,k\}$. We simulate $m=100$ observations for ${\bm\mu}=\mu_{1:4}=(-3,0,3,6)$. The resulting posterior distribution for ${\bm\mu}$ is given by
\[
p({\bm\mu}|\by)\propto p(\by|{\bm\mu})\ind({\bm\mu}\in\lbrack-10,10]^{4})
\]
The main computational challenge associated to Bayesian inference in finite mixture models is the nonidentifiability of the components. As we have used exchangeable priors for the parameters $\mu_{1:4}$, the posterior distribution $p({\bm\mu}|\by)$ is invariant to permutations in the labelling of the parameters. Hence this posterior admits $k!=24$ symmetric modes.

Generating $N$ samples from such a posterior is a popular method for determining the ability of samplers to explore a high-dimensional space with multiple well-separated modes, which should all be represented in the samples. Basic random-walk MCMC and importance sampling methods typically fail to provide a correct approximation of the posterior for practical values of $N$ \citep{celeux}. It should be noted that while it might not be necessary to sample from all the symmetric modes in the case of a mixture model, the successful traversal of all the modes suggests that the sampler would succeed in traversing non-symmetric modes in other distributions, so long as symmetry is not exploited by the sampler.

\subsubsection{Population-Based MCMC}
\label{section:pb_mcmc}
We select the auxiliary distributions $\pi_{1:M-1}$ following the parallel tempering methodology, i.e. $\pi_{i}(\bx) \propto\pi(\bx)^{\beta_{i}}$ with $0 < \beta_{1} < \cdots< \beta_{M} = 1$ and use $M = 200$. This class of auxiliary distributions is motivated by the fact that MCMC converges more rapidly when the target distribution is flatter. For this problem, we use the cooling schedule $\beta_{i} = (i/M)^{2}$ and a standard $\N({\bm 0},I_{k})$ random walk Metropolis-Hastings kernel for the first stage moves.

For the second stage moves, we use only the basic exchange move \citep{geyer,hukushima_nemoto}: chains $i$ and $j$ swap their values with probability $\min\{1,\alpha_{ij}\}$ where
\[
\alpha_{ij}=\frac{\pi_{i}(\bx_{j})\pi_{j}(\bx_{i})}{\pi_{i}(\bx_{i})\pi_{j}(\bx_{j})}
\]
Further, we allow exchanges to take place only between adjacent chains so that all moves can be done in parallel. We use $R=M/2$ and $\I_{1:R}$ is either $\{\{1,2\},\{3,4\},\ldots,\{M-1,M\}\}$ or $\{\{2,3\},\{4,5\},\ldots,\{M-2,M-1\},\{M,1\}\}$, each with probability half. While use of permutation or crossover moves would be appropriate for this particular model, we felt that they would detract from the ability to generalize our results to the case where the likelihood is not invariant to permutations of the labels.

\begin{table}[htp]
\small
\caption{Running times for the Population-Based MCMC Sampler for various numbers of chains $M$. \vspace{10pt}}
\label{tab:pop_mcmc}
\centering
\begin{tabular}{| c | c | c | c | c | c |}
\hline
$M$            & CPU (mins)    & 8800GT (secs)   & Speedup      & GTX280  (secs) & Speedup\\
\hline
8              & 0.0166        & 0.887           & 1.1          & 1.083       & 0.9     \\
\hline
32             & 0.0656        & 0.904           & 4            & 1.098       & 4     \\
\hline
128            & 0.262         & 0.923           & 17           & 1.100       & 14    \\
\hline
512            & 1.04          & 1.041           & 60           & 1.235       & 51    \\
\hline
2048           & 4.16          & 1.485           & 168          & 1.427       & 175   \\
\hline
8192           & 16.64         & 4.325           & 230          & 2.323       & 430   \\
\hline
32768          & 66.7          & 14.957          & 268          & 7.729       & 527   \\
\hline
131072         & 270.3         & 58.226          & 279          & 28.349      & 572   \\
\hline
\end{tabular}
\end{table}

\begin{figure}[pth]
\center
\includegraphics[scale=0.5]{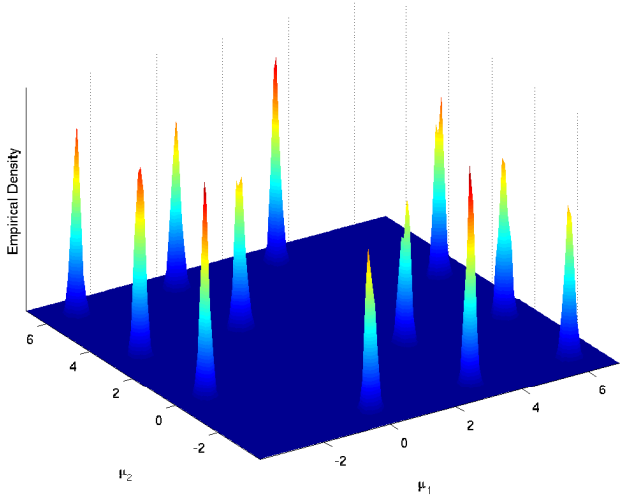}
\caption{Estimated marginal posterior density $p(\mu_{1:2}|\by)$ from MCMC samples}
\label{fig:modes_12_mcmc}
\end{figure}

\begin{figure}[pth]
\center
\includegraphics[scale=0.5]{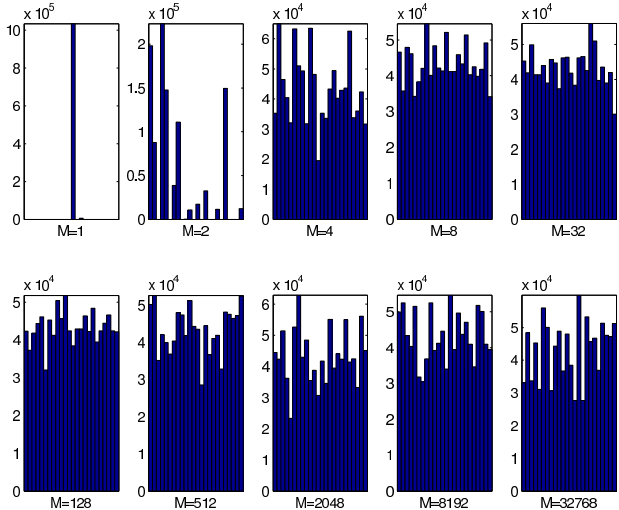}
\caption{Number of MCMC samples from each mode}
\label{fig:bar_modes_mcmc}
\end{figure}

To test the computational time required by our algorithms we allow the number of chains to vary but fix the number of points we wish to sample from the marginal density $\pi_{M}=\pi$ at 8192. As such, an increase in the number of chains leads to a proportional increase in the total number of points sampled. Processing times for our code are given in Table \ref{tab:pop_mcmc}, in which one can see that using $131072$ chains is impractical on the CPU but entirely reasonable using the GPU. Figure \ref{fig:modes_12_mcmc} shows the estimated posterior density $p(\mu _{1:2}|\by)$ from a set of $2^{20}$ MCMC samples from $\pi_{M}$ with $M=32768$, which is nearly identical to the estimated marginal posterior densities of any other pair of components of ${\bm\mu}$. This marginal density has 12 well-separated modes in $\Re^{2}$ but it is worth noting that the joint density $p(\mu_{1:4}|\by)$ has 24 well-separated modes in $\Re^{4}$. Figure \ref{fig:bar_modes_mcmc} shows the number of points from each mode for various values of $M$. We also computed the average number of iterations taken for the samplers to traverse all modes for the different values of $M$. For $M = 1$ and $M = 2$, the sampler did not traverse all the modes at all, while for values of $M$ between 4 and 32 the traversal time decreased from 80000 to 10000, after which it was unchanged with increases in $M$. These numbers should be compared to $24\times H_{24} \approx 91$ - the expected number of samples required to cover every mode if one could sample independently from $\pi$ - where $H_{i}$ is the $i$th harmonic number.

\subsubsection{SMC Sampler}
As with population-based MCMC, we use a tempering approach and the same cooling schedule, i.e. $\pi_{t}(\bx)\propto\pi(\bx)^{\beta_{t}}$ with $\beta_{t}=(t/M)^{2}$ and $M=200$. we use the uniform prior on the hypercube to generate the samples $\{\bx_{0}^{(1:N)}\}$ and perform 10 MCMC steps with the standard $\N({\bm0},I_{k})$ random walk Metropolis-Hastings kernel at every time step. We use the generic backwards kernel suggested in \cite{crooks}, \cite{ais} and \cite{smcs} for the case where each kernel is $\pi_{t}$-stationary so that the unnormalized incremental importance weights are of the form $\pi_{t}(\bx_{t-1})/\pi_{t-1}(\bx_{t-1})$. 
\begin{table}[htp]
\small
\caption{Running times for the Sequential Monte Carlo Sampler for various values of $N$. \vspace{10pt}}
\label{tab:smcs}
\centering
\begin{tabular}{| c | c | c | c | c | c |}
\hline
$N$            & CPU (mins)    & 8800GT (secs)   & Speedup      & GTX280  (secs) & Speedup\\
\hline
8192           & 4.44          & 1.192           & 223.5        & 0.597          & 446   \\
\hline
16384          & 8.82          & 2.127           & 249          & 1.114          & 475   \\
\hline
32768          & 17.7          & 3.995           & 266          & 2.114          & 502   \\
\hline
65536          & 35.3          & 7.889           & 268          & 4.270          & 496   \\
\hline
131072         & 70.6          & 15.671          & 270          & 8.075          & 525   \\
\hline
262144         & 141           & 31.218          & 271          & 16.219         & 522      \\
\hline
\end{tabular}
\end{table}

\begin{figure}[pth]
\center
\includegraphics[scale=0.5]{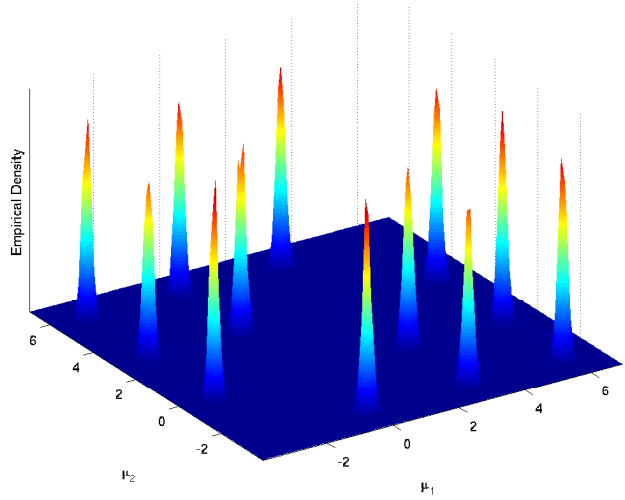}
\caption{Estimated marginal posterior density $p(\mu_{1:2}|\by)$ from SMC samples}
\label{fig:modes_12_smcs}
\end{figure}

\begin{figure}[pth]
\center
\includegraphics[scale=0.5]{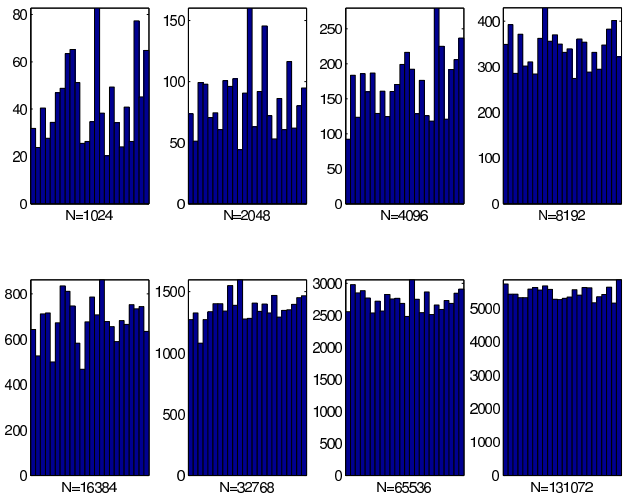}
\caption{Effective number of SMC samples from each mode}
\label{fig:bar_modes_smcs}
\end{figure}

We also ran the SMC sampler with no resampling at all, which for these settings corresponds to the Annealed importance sampling (AIS) method proposed in \cite{ais}; see also \cite{crooks} for a similar method in physics. The resulting samples were less successful at characterizing the full multi-modality of the posterior distribution. This is consistent with the numerical findings and theoretical results discussed in \cite[Sections 3.5 and 4.2.3]{smcs}: when the MCMC\ kernels used at every time step mixes reasonably well, it is very beneficial to resample. In addition, when ESS is used as a threshold for resampling in the SMC sampler, the resampling step takes very little time compared to the evolution and weighting of the particles simply because it happens so infrequently. As such, the running time of the SMC sampler compared to AIS is practically identical. Processing times for our code are given in Table \ref{tab:smcs}. Figure \ref{fig:modes_12_smcs} shows the estimated posterior density $p(\mu_{1:2}|\by)$ from the SMC sampler with $N=65536$. Figure \ref{fig:bar_modes_smcs} shows the number of points from each mode for various values of $N$.

\subsubsection{Comparison}
\label{section:comparison}
While both methods are capable of exploring the posterior distribution for ${\bm \mu}$, there are important differences in how the methods make use of parallelization. In particular, the SMC sampler parallelizes across particles approximating the same auxiliary distribution whilst the MCMC sampler parallelizes across auxiliary distributions at the same iteration. As such, to make full use of the graphics card the SMC sampler requires many particles while the MCMC sampler requires many auxiliary distributions. In most cases, however, one will be happy to use in excess of 8192 particles for SMC but one may not want to use in excess of 32768 auxiliary distributions. Indeed, for the application described above there seems to be no benefit in increasing the number of chains beyond 128, although this might also be due to the choice of cooling schedule and random walk variances. Furthermore, we utilized only the simplest information exchange and proposal moves in our samplers so as not to trivialize the problem. It should be noted, however, that there are situations in which a large number of intermediate temperatures are required for exchange acceptance probabilities to be greater than some preset value, for example when the dimension of the distribution of interest increases \cite{predescu}. As mentioned in Section \ref{section:pb_mcmc}, we would like to emphasize that the use of very large numbers of chains is possible using these parallel methods with only a modest increase in computation time.

The SMC sampler appears to be more efficient than the MCMC sampler for this problem. Indeed, with only 8192 particles the SMC sampler gives a reasonable representation of the posterior, taking only 597ms. The MCMC sampler requires around $2^{20}$ samples to give a reasonably uniform number of samples per mode, and this takes just over 2 minutes. In addition, although we have not done so here, it is possible with both the SMC and MCMC approaches to use the samples from the auxiliary distributions to estimate integrals of interest by computing appropriate importance weights. An interesting recent proposal on how to effectively combine estimates using such samples can be found in \cite{gramacy}.

For Bayesian inference in mixture models, there are many ways of dealing with the identifiability of the mixture parameters; \cite{jasra_thesis} includes a review of these. It is worth mentioning that for this type of model, we can permute samples as a post-processing step or within an MCMC kernel so traversal of the modes can be achieved trivially. The speedup of both methods is unaffected by the use of such mechanisms. In addition, the speedup is unaffected by increases in the number of observations since this only increases the amount of computation that each thread must do by a constant factor. Increasing the number of observations also has little effect on the difficulty for the sampler to move between modes since the modes are already well separated. Similarly, the speedup observed is robust to changes in the number of mixture components. The computation of each likelihood requires memory that is linear in the number of components, while the memory required per thread dictates the number of threads that can be run in parallel. However, as the number of components increases we usually have more observations, providing an opportunity to parallelize the computation of a single likelihood across multiple threads. This allows the amount of memory per thread to be significantly lower than if the complete likelihood was calculated by each thread and thus still allows many threads to be run in parallel.

\subsection{Factor Stochastic Volatility}
\label{section:fsv}
Many financial time series exhibit changing variance. A simple multivariate volatility model that allows us to capture the changing cross-covariance patterns of time series consists of using a dynamic latent factor model. In such models, all the variances and covariances are modelled through a low dimensional stochastic volatility structure driven by common factors \citep{fsv_liu,tvc_fsv}. We consider here a factor stochastic volatility model most similar to that proposed in \cite{fsv_liu}:
\begin{align*}
\by_{t} &  \sim\N(\mathbf{B}\mathbf{f}_{t},{\bm\Psi})\\
\mathbf{f}_{t} &  \sim\N(\mathbf{0},\mathbf{H}_{t})\\
{\bx}_{t} &  \sim\N({\bm\Phi}{\bx}_{t-1},\mathbf{U})
\end{align*}
where
\begin{align*}
{\bm\Psi} &  \eqdef\text{diag}(\psi_{1},\ldots,\psi_{M})\\
\mathbf{H}_{t} &  \eqdef\text{diag}(\exp({\bx}_{t}))\\
{\bm\Phi} &  \eqdef\text{diag}(\phi_{1},\ldots,\phi_{K})
\end{align*}

Here, $\mathbf{f}_{t}$ is $K$-dimensional, $\by_{t}$ is $M$-dimensional and $\mathbf{B}$ is an $M \times K$ factor loading matrix with zero entries above the diagonal for reasons of identifiability. The latent variable at each time step $t$ is the $K$-dimensional vector ${\bx}_{t}$. The likelihood of the data, $\by_{t}$, given ${\bx}_{t}$ is Gaussian with
\begin{align*}
\by_{t} | {\bx}_{t} \sim\N(\mathbf{0},\mathbf{B} \mathbf{H}_{t}
\mathbf{B}^{T} + {\bm \Psi})
\end{align*}

We generate data for times $t = 1,\ldots,T = 200$, $M = 5$, $K = 3$, ${\bx}_{0} = \mathbf{0},\psi_{i} = 0.5, i \in\{1,\ldots,M\}$, $\phi_{i} = 0.9, i \in\{1,\ldots,K\}$,
\[
\mathbf{B} = \left(
\begin{array}
[c]{ccc}%
1 & 0 & 0\\
0.5 & 1 & 0\\
0.5 & 0.5 & 1\\
0.2 & 0.6 & 0.3\\
0.8 & 0.7 & 0.5
\end{array}
\right)  \text{ and } \mathbf{U} = \left(
\begin{array}
[c]{ccc}%
0.5 & 0.2 & 0.1\\
0.2 & 0.5 & 0.2\\
0.1 & 0.2 & 0.5
\end{array}
\right)
\]

\begin{figure}[pth]
\center
\includegraphics[scale=0.5]{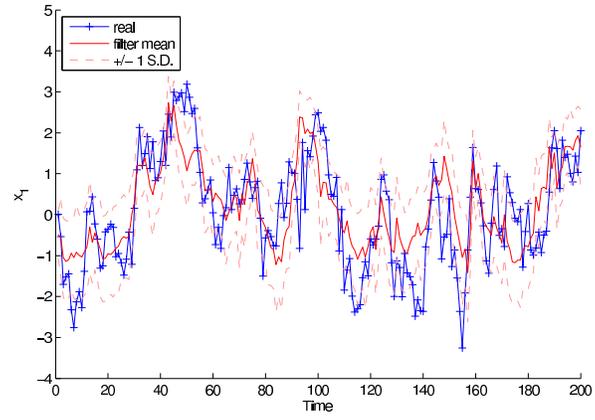}
\includegraphics[scale=0.5]{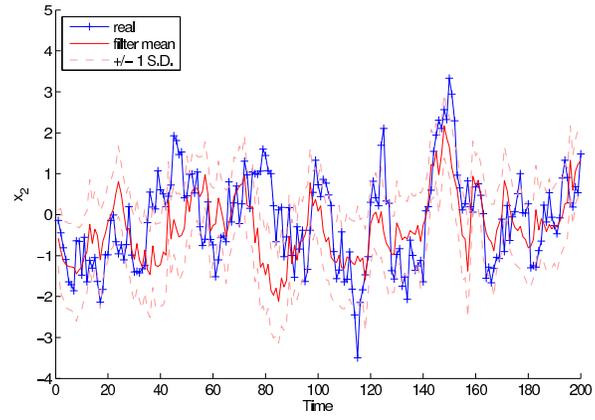}
\includegraphics[scale=0.5]{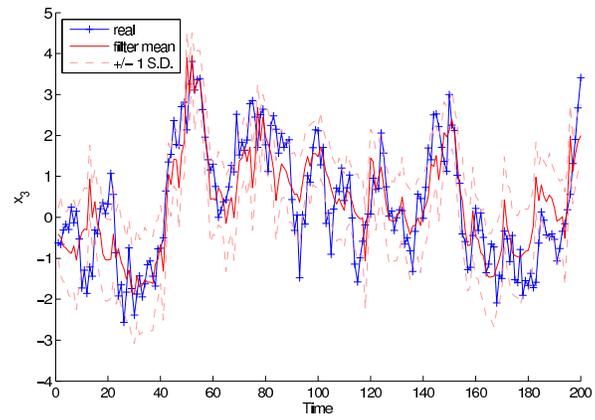}
\caption{Estimated and real values of $\bx$}
\label{fig:fsv_alpha}
\end{figure}

This is a simple example of a multivariate, non-linear and non-Gaussian continuous state-space model for which particle filters are commonly employed to sample from the posterior $p({\bx}_{0:T}|\by_{1:T})$. Processing times for our code are given in Table \ref{tab:smc}. In Figure \ref{fig:fsv_alpha} we plot the filter means for each component of ${\bx}$ with $\pm1$ sample standard deviations alongside the true values of ${\bx}$ used to generate the observations.

\begin{table}[htp]
\small
\caption{Running time (in seconds) for the Sequential Monte Carlo method for various values of $N$. \vspace{10pt}}
\label{tab:smc}
\centering
\begin{tabular}{| c | c | c | c | c | c |}
\hline
$N$            & CPU           & 8800GT          & Speedup      & GTX280     & Speedup \\
\hline
8192           & 2.167         & 0.263           & 8            & 0.082      & 26      \\
\hline
16384          & 4.325         & 0.493           & 9            & 0.144      & 30      \\
\hline
32768          & 8.543         & 0.921           & 9            & 0.249      & 34      \\
\hline
65536          & 17.425        & 1.775           & 10           & 0.465      & 37      \\
\hline
131072         & 34.8          & 3.486           & 10           & .929       & 37      \\
\hline
\end{tabular}
\end{table}

The speedups obtained in this application are considerably less than for mixture model inference problem. This can be explained by lower arithmetic intensity, higher space complexity in each thread and increased resampling rate as compared to the SMC sampler example above. The mixture model likelihood calculation contains a compute-intensive product-sum operation involving 104 values whilst the factor stochastic volatility likelihood consists mainly of matrix operations. In the latter case, the speedup is independent of $T$ but not the dimension of the observations since the amount of memory required per thread increases quadratically in the dimension of each observation. For example, we attained a speedup of 80 on the GTX 280 when running a particle filter for a multivariate stochastic volatility model with $M = K = 2$. The frequency of resampling is an issue with respect to speedup because it can typically only attain around 10 to 20 fold speedup for practical values of $N$, mainly due to the parallel scan operation. This potentially gives rise to tradeoffs in speedup between the transition and weighting steps and the time between resampling steps for some models, since more sophisticated proposal distributions that parallelize less cleanly might reduce the resampling rate. This type of performance, however, still provides considerable speedup and may be more representative of the type of speedup practitioners can expect in general.

\subsection{Floating Point Precision}
For all three algorithms discussed above, we ran identical algorithms with the same random numbers on the CPU using double precision floating point numbers and the resulting estimates of expectations of interest were affected by an order of magnitude less than the Monte Carlo variance of the estimates. The actual paths sampled, of course, were different due to the sensitivity of all of the algorithms to perturbations but this did not affect the ability of the samples to approximate the target distribution.

\section{Discussion}
\label{section:discussion}
The speedup for the population-based MCMC algorithm and the SMC sampler is tremendous. In particular, the evaluation of $p(\by|{\bm \mu})$ for the mixture-modelling application has high arithmetic intensity since it consists of a product-sum operation with 400 Gaussian log-likelihood evaluations involving only 104 values. In fact, because of the low register and memory requirements, so many threads can be run concurrently that SIMD calculation of this likelihood can be sped up by 500 times on the 8800 GT and 800 times on the GTX 280. However, the speedup attained for the standard SMC algorithm may be more representative of the kinds of gains one can expect in most applications with only reasonable arithmetic intensity. Even so, speedups of 10 to 35 make many problems tractable that previously were not by reducing a week's worth of computation to a few hours. For example, estimation of static parameters in continuous state-space models or the use of SMC proposals within MCMC can require thousands of runs, so a speedup of this scale can substantially reduce the computation time of such approaches (see e.g. \cite{pmcmc}). It is worth noting also that we can expect speedups in the vicinity of 500 with SMC if few resampling steps are required and each weighting step has small space complexity and moderate time complexity.

The GTX 280 GPU outperforms the 8800 GT GPU by a factor of about 2 in all situations in which the GPU is used to capacity. This is the case in all but the population-based MCMC algorithm, in which the number of threads is determined by the number of auxiliary distributions. The reason for this is simple: the algorithms presented are register-bound on the inputs given, in that the number of registers required by each thread is the critical quantity that bounds the number of threads that can be run concurrently. The GTX 280 has twice the number of registers per multiprocessor and more than twice the multiprocessors compared to the 8800 GT. Hence, one could expect more speedup on many-core chips with even more registers. In fact, further improvements could be made using multiple cards with large amounts of memory, configurations of which are now available in NVIDIA's Tesla line. These Tesla `personal supercomputers' comprise 3 or more high-performance GPUs, each with 4GB of memory and a CPU with at least as much memory as the GPUs' combined.  It is also possible to design algorithms that are memory-bound, though we have not encountered this in the context of Monte Carlo simulation. It is certainly possible that both register and memory limitations can affect the parallelizability of the mentioned algorithms when facing very high-dimensional problems. However, in such cases it is possible that alternative uses of many-core architecture can provide speedup in these situations.

The acceleration of the Monte Carlo methods discussed here have practical benefits not only to computation time but also to energy efficiency. A general purpose CPU allocates extra circuits and hence power to flow control and caching which is unnecessary for the types of computation described here. As such, reasonable decreases in power consumption can be realized by using specialized many-core architectures like SIMD instead.

It should be noted that while we have used CUDA to implement the parallel components of algorithms, the results are not necessarily specific to this framework or to GPUs. It is expected that the many-core processor market will grow and there will be a variety of different devices and architectures to take advantage of. However, the SIMD parallelization technique and the sacrifice of caching and flow control for arithmetic processing is unlikely to disappear, particularly because when it is well-suited to a problem it will nearly always deliver considerable speedup. In addition, GPUs are affordable, off-the-shelf components that can be easily installed on a personal computer. Of particular interest is an emerging framework, the Open Computing Language (OpenCL), which provides a uniform programming environment for developers that enables them to write portable code for a variety of parallel devices, including GPUs and CPUs. For users who would like to see moderate speedup with very little effort, there is work being done to develop libraries that will take existing code and automatically generate code that will run on a GPU. An example of this is the Jacket engine for MATLAB code, created by Accelereyes.

The speedups attainable with many-core architectures have broad implications in the design, analysis and application of SMC and population-based MCMC methods. With respect to SMC, it allows more particles to be used for the same or even less computation time, which can make these samplers viable where they previously were not. When faced with designing a population-based MCMC sampler, the results expectedly show that there is little cost associated with increasing the number of auxiliary distributions until the GPU reaches the critical limit of threads it can run concurrently. After this, there is a doubling in the computation time when the number of chains is doubled. In our application, this does not occur until we have around 4096 auxiliary distributions. One might notice that this number is far larger than the number of processors on the GPU. This is due to the fact that even with many processors, significant speedup can be attained by having a full pipeline of instructions on each processor to hide the relatively slow memory reads and writes. Of course, we can expect this application-specific number to decrease when dealing with higher-dimensional distributions or those whose density evaluations require more registers or memory. Nevertheless, practitioners have more freedom to increase the number of auxiliary distributions to achieve a faster rate of convergence as the computation time associated with each step is not as closely tied to this value as it is on a single-core processor. In both SMC and MCMC, it is also clear from this case-study that it is beneficial for each thread to use as few registers as possible since this determines the number of threads that can be run simultaneously. This may be of interest to the methodology community since it creates a space-time tradeoff that might be exploited in some applications.

A consequence of the space-time tradeoff mentioned above is that methods which require large numbers of registers per thread are not necessarily suitable for parallelization using GPUs. For example, operations on large, dense matrices that are unique to each thread can restrict the number of threads that can run in parallel and hence dramatically affect potential speedup. In cases where data is shared across threads, however, this is not an issue. For example, a mixture model with large amounts of data does not affect the number of registers required whilst increasing the number of components increases the number of registers required only linearly. In contrast, increasing the number of observed assets in a factor stochastic volatility model leads to a quadratic increase in the number of registers required, substantially affecting scalability in this regard. An increase in the number of observations itself has no effect on speedup. In principle, it is not the size of the data that matters but the space complexity of the algorithm in each thread that dictates how scalable the parallelization is.

The parallelization of the advanced Monte Carlo methods described here opens up challenges for both practitioners and for algorithm designers. There are already an abundance of statistical problems that are being solved computationally and technological advances, if taken advantage of by the community, can serve to make previously impractical solutions eminently reasonable and motivate the development of new methods.

\section*{Acknowledgments}
The authors acknowledge support from the Oxford-Man Institute for Quantitative Finance and the Medical Research Council. Anthony Lee is additionally funded by a Clarendon Fund Scholarship and Christopher Yau is funded by a UK Medical Research Council Specialist Training Fellowship in Biomedical Informatics (Ref No. G0701810).

\appendix
\section*{Appendix: Web Resource}
We have created a website resource at \url{http://www.oxford-man.ox.ac.uk/gpuss/} for the statistics community with the code used in these examples as well as useful information on GPU programming for statistical computing with links to tutorials and relevant papers.

\bibliography{cuda_mc_arxiv_v3}

\end{document}